*Gene expression*

# Elucidation of Directionality for Co-Expressed Genes: Predicting Intra-Operon Termination Sites

Anshuman Gupta[1], Costas D. Maranas[2], Réka Albert[3,*]

[1]Academic Services and Emerging Technologies, [2]Department of Chemical Engineering and [3]Department of Physics, The Pennsylvania State University, University Park, PA, USA

**ABSTRACT**

**Motivation:** In this paper, we present a novel framework for inferring regulatory and sequence-level information from gene co-expression networks. The key idea of our methodology is the systematic integration of network inference and network topological analysis approaches for uncovering biological insights.

**Results:** We determine the gene co-expression network of *Bacillus subtilis* using Affymetrix GeneChip® time series data and show how the inferred network topology can be linked to sequence-level information hard-wired in the organism's genome. We propose a systematic way for determining the correlation threshold at which two genes are assessed to be co-expressed by using the *clustering coefficient* and we expand the scope of the gene co-expression network by proposing the *slope ratio* metric as a means for incorporating directionality on the edges. We show through specific examples for *B. subtilis* that by incorporating expression level information in addition to the temporal expression patterns, we can uncover sequence-level biological insights. In particular, we are able to identify a number of cases where (*i*) the co-expressed genes are part of a single transcriptional unit or *operon* and (*ii*) the inferred directionality arises due to the presence of intra-operon transcription termination sites.

**Availability:** Provided on request.

**Contact:** Professor Réka Albert. Phone: 814-865-6123. E-mail: ralbert@phys.psu.edu

## 1   INTRODUCTION

Gene expression information captured in microarray data for a variety of environmental and genetic perturbations, in conjunction with other sources such as protein-protein/protein-DNA interaction and operon organization data, promises to yield unprecedented insights into the organization and functioning of biological systems (Brazhnik *et al.*, 2002; Ge *et al.*, 2003). It has now become clear that only by simultaneously accounting for all the multiple layers of regulation that exist in biological systems can one hope to uncover the biological *knowledge* embedded in the experimental *data*. In this paper, we take a first step by proposing an integrated approach that combines network inference and analysis with a detailed biological study of the uncovered regulatory patterns. Specifically, we study the gene co-expression network of *Bacillus subtilis* derived from Affymetrix GeneChip® time series data and show how the observed expression patterns can be traced back to the sequence-level organization of the various genes. We construct a gene co-expression network where nodes correspond to genes and an undirected edge exists between two genes if the similarity of their expression profiles exceeds a threshold value. We introduce a novel systematic way for determining the similarity threshold by applying ideas from network analysis. In addition, we expand the scope of the gene co-expression network representation by proposing a way to impose directionality on the edges. We show that this enables us to uncover additional biologically relevant insights that are not obvious at the level of the gene co-expression network.

Several methods aim to cluster genes on the basis of their expression profiles for identifying groups of genes that are co-expressed/co-regulated under particular experimental conditions (Eisen *et al.*, 1998; Wen *et al.*, 1998; Herwig *et al.*, 1999; Tamayo *et al.*, 1999; Dougherty *et al.*, 2002; Xu *et al.*, 2002; Schmitt *et al.*, 2004). The graph-based method that we use is closely related to the widely used hierarchical clustering approach (Eisen *et al.*, 1998; Spellman *et al.*, 1998). In hierarchical clustering, one starts with all genes belonging to separate clusters. Subsequently, the two genes which are closest to each other with respect to a chosen similarity measure are assigned to a single cluster. A similarity measure between gene clusters is then used to determine the most similar pair of clusters, which are then merged. This procedure is repeated until all the genes belong to a single cluster and the clustering results are displayed as a dendrogram. Specific clusters are obtained by "slicing" the dendrogram at a threshold of the similarity measure. In the same spirit, in our proposed approach all genes belong to a single cluster when the similarity threshold is set to zero while they constitute individual single-gene clusters when the similarity threshold is set at the other extreme of one. A common question to be answered for both hierarchical clustering and our graph-based approach is where to set the similarity threshold. In this work, we propose an objective and systematic method of determining the most informative similarity threshold by employing ideas from network analysis. Another key advantage of our method over hierarchical clustering is that we do not need to define a similarity measure between clusters and can rely solely on the similarity measure between individual genes for obtaining the clusters.

---

[*]To whom correspondence should be addressed.

## 2 METHODS

**Gene Co-Expression Network**

The starting point of our analysis is time-series expression information for 746 *B. subtilis* genes in a cradle-to-grave experiment (Gupta *et al.*, 2005). *B. subtilis* is one of the best characterized gram positive bacteria and it is used extensively for industrial-scale protein production. The expression data were collected during feed-batch protease production and the majority of the genes included were involved in the central metabolism of *B. subtilis*. A total of 20 time points were sampled every 2 hours over the course of the fermentation. Since the data was obtained using Affymetrix arrays, the expression levels were time-resolved absolute transcript signals as opposed to the relative expression changes typically measured with cDNA arrays.

In this work, we use the R-square metric (or equivalently the Pearson correlation coefficient) as a measure of similarity between the expression profiles of two genes. This metric, which ranges between 0 and 1, quantifies the "goodness-of-fit" of a linear relationship between two variables. Suppose that the system is described by two variables $X$ and $Y$ where $Y$ is linearly dependent on $X$ i.e., $Y = a_{YX} + b_{YX} X$. Given observed data $(Y_i, X_i)$, the parameters $a_{YX}$ and $b_{YX}$ are determined by solving the following quadratic least-squares minimization problem (Rencher, 1995).

$$\min_{a_{YX}, b_{YX}} SSE_{YX} = \sum_{i=1}^{N} (Y_i - a_{YX} - b_{YX} X_i)^2$$

The optimal regression slope, which is determined by setting the first derivatives $\partial SSE_{YX}/\partial a_{YX}$ and $\partial SSE_{YX}/\partial b_{YX}$ to zero and solving the resulting 2x2 system of linear equations, is given by

$$b_{YX} = \frac{\sum_{i=1}^{N}(X_i - \overline{X})\cdot(Y_i - \overline{Y})}{\sum_{i=1}^{N}(X_i - \overline{X})^2}$$

where

$$\overline{X} = \frac{1}{N}\sum_{i=1}^{N} X_i \; ; \; \overline{Y} = \frac{1}{N}\sum_{i=1}^{N} Y_i$$

Alternatively, we can regress "X-on-Y" by switching the roles of the independent and dependent variables and determine the alternative regression slope $b_{XY}$. If the relationship between $X$ and $Y$ is *perfectly* linear, then the two slopes are reciprocal of each other, i.e., $b_{XY} \cdot b_{YX} = 1$. To measure the degree of linearity, the $R^2$ metric is defined as $R^2 = b_{XY} \cdot b_{YX}$. In our current setting where we are given time course gene expression data $X_{it}$ for each gene $i$ at time point $t$, $R^2_{ij}$ is given by

$$R^2_{ij} = \frac{\left[\sum_{t=1}^{T}(X_{it} - \overline{X}_i)\cdot(X_{jt} - \overline{X}_j)\right]^2}{\sum_{t=1}^{T}(X_{it} - \overline{X}_i)^2 \cdot \sum_{t=1}^{T}(X_{jt} - \overline{X}_j)^2}$$

Next, the co-expression network is defined such that each gene corresponds to a node and an undirected edge is included between two nodes if the R-square value between the two genes is greater than a threshold value $R^*$. Thus, the connectivity matrix $a_{ij}$, which encodes whether a particular gene/node $i$ is connected to another gene/node $j$, is given by

$$a_{ij} = 1 \quad if \quad R^2_{ij} \geq R^* \quad ; \quad a_{ij} = 0 \quad if \quad R^2_{ij} < R^*$$

where $R^2_{ij}$ is the R-square value between the time-courses of genes $i$ and $j$. Traditionally the similarity threshold $R^*$ has been chosen either by assuming a normal R-square distribution (which may or may not be valid) or by significance analysis using the R-square null distribution generated by permutation of the original data (Magwene and Kim, 2004). Though valid from the perspective of trying to minimize the number of false-positive edges, both these methods fail to take into consideration the linearity assumptions that underlie the R-square metric and do not provide any information regarding the sensitivity of the resulting network topology to the chosen cut-off value. In this work we determine the similarity threshold by employing a commonly used graph-theoretic transitivity measure, the clustering coefficient.

**Clustering Coefficient**

The clustering coefficient $C_i$ of a node (Watts and Strogatz, 1998; Albert and Barabasi, 2002) is defined as

$$C_i = \frac{E_i}{k_i(k_i - 1)/2}$$

where $k_i$ (>1) is the number of first-neighbors of node $i$ (*i.e.*, nodes that are connected to it by a single edge) and $E_i$ is the number of edges present between these first neighbors. If the $k_i$ first-neighbors were all connected to each other, then there would be a total of $k_i(k_i - 1)/2$ edges between them, leading to a clustering coefficient of 1 for node $i$.

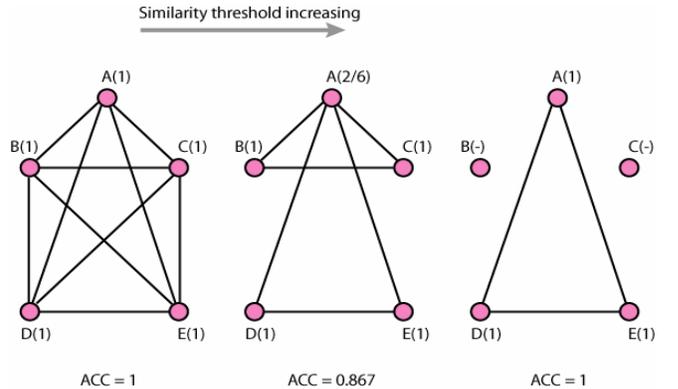

**Fig. 1**. Impact of the similarity threshold on the average clustering coefficient of a network. The clustering coefficient of each node is shown next to the node label (ACC = Average Clustering Coefficient). Note that in the third graph the clustering coefficient is not defined for nodes B and C and so they are not considered for calculating the ACC.

Figure 1 shows how the similarity threshold impacts the average clustering coefficient of an example network. If the similarity threshold is set at its minimum value of zero, a fully connected network with an average clustering coefficient (ACC) of one will be obtained. As the cut-off is increased, edges will get eliminated leading to a lowering of the ACC. However, at a reasonably high similarity threshold, one can expect the ACC to start increasing once again due to the emergence of highly "cliquish" disconnected sub-networks or modules, a manifestation of the transitive property of linear functions at the network representation level. This predictable variation of the ACC with the similarity threshold suggests that in the general case, the similarity threshold should be chosen such that (*i*) it is low enough so that a sufficient number of relationships (edges) are retained in the first place, (*ii*) it is consistent with the transitive property of linear relationships which is what it eventually captures and finally (*iii*) it is high enough so that the probability of the implied linear relationships (edges) occurring by chance was low. Thus, by parametrically varying the similarity threshold and calculating the resulting ACC, the "critical" threshold should be set at the highest similarity value above/below which a sharp increase/decrease in the ACC is observed.

**Edge Directionality**

A high similarity between gene expression profiles could mask important differences in the regulation of the gene expression levels. To highlight this point, consider two representative gene pairs shown in Figure 2. Even though the R-square values are comparable for the two gene pairs, it is clear from their expression profiles that there is a qualitative difference between the two pairs. Specifically, for genes A and B, not only is the overall *shape* of the expression profile similar, but also the *magnitudes* of the expression levels are almost identical at each of the time points. However, the level of peak expression of gene D is much higher than that of gene C. In view of this, we considered the regression slopes for the two gene pairs $b_{AB} = 1.004$, $b_{BA} = 0.976$, $b_{CD} = 0.198$, $b_{DC} = 4.957$. To quantify how different these slopes were we defined the *slope ratio* metric $SR$ as

$$SR = \frac{\min(|b_{YX}|, |b_{XY}|)}{\max(|b_{YX}|, |b_{XY}|)}$$

This metric allows for distinguishing between the two pairs since $SR_{CD} = 0.04$ is significantly different than $SR_{AB} = 0.97$. We incorporat the slope ratio information at the network representation level by assigning directionality to *only* those edges that have $SR \to 0$ according to the following rules.

$$If \quad SR = \frac{|b_{YX}|}{|b_{XY}|} \Rightarrow Y \to X \quad ; \quad If \quad SR = \frac{|b_{XY}|}{|b_{YX}|} \Rightarrow X \to Y$$

The interpretation of the directed edge between two genes (for example, $C \to D$) is that a small change in the "source" gene $(C)$ is associated with a large change in the "sink" gene $(D)$. Consequently, by using the slope ratio metric in tandem with the R-square metric, we can identify *differentially* (in terms of absolute level) *co-expressed* (in terms of temporal pattern) gene pairs such as C and D.

In addition to the similarity threshold, we will now need to determine a SR threshold such that directed edges having SR value less than the threshold are considered significant. We use the distribution of SR values for the un-directed edges surviving the similarity threshold for determining the SR threshold. Specifically, in keeping with the spirit of how the similarity threshold is determined, the SR threshold is chosen as the critical value around which a sharp transition is observed in the SR distribution. By doing so we are able to focus on gene pairs that "stand out" from the background of all co-expressed gene pairs.

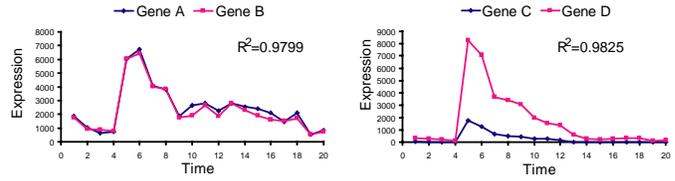

**Fig. 2.** Two qualitatively different gene pairs that cannot be differentiated purely on the basis of the R-square metric

## 3 RESULTS

Figure 3 highlights the dependence of the average clustering coefficient (ACC) of the co-expression network on the R-square threshold. Note that we excluded both the isolated nodes (node degree 0) and the leaf nodes (node degree 1) from the calculation of the ACC. As expected, when the threshold was set at 0, resulting in a fully connected network (since the R-square is greater than 0 for *all* gene pairs), an ACC of 1 was obtained. A small increase in the cut-off from 0 resulted in a sharp decline in the ACC as a result of edge deletion from the fully connected network. The ACC leveled off around 0.65 for a broad range of cut-off values (0.12 to 0.88). From a network topology perspective, this implied that the average cohesiveness of the neighborhood of the non-isolated nodes was largely constant even though the number of edges was continuously decreasing over this range. On subsequent increase of the threshold (0.9 and above) a sharp increase in the ACC was observed. This was, as expected, due to the emergence of highly self-connected sub-networks or modules. Given this plot, we chose the similarity threshold as 0.90 since we observe a sharp transition in the ACC around this value.

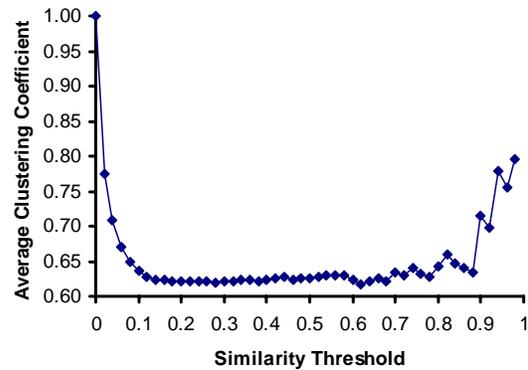

**Fig. 3**. Variation of the average clustering coefficient with the similarity threshold for the 746 gene "cradle-to-grave" data-set.

The resulting co-expression network is shown in Figure S1 of the Supplementary Data. There were 47 separate clusters (sub-networks) in the network with the largest connected component consisting of 79 nodes and 260 edges (see Table S1 in Supplementary Data). Figure 4 shows the characteristic expression signatures of the top 6 clusters. These clusters were mainly composed of genes involved in t-RNA synthesis, amino acid metabolism (in particular tryptophan), biotin synthesis and pentose and glucuronate interconversion pathway (see Supplementary Data).

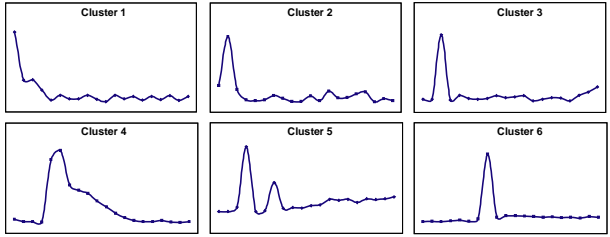

**Fig. 4.** Expression signatures for the top 6 clusters in the gene co-expression network.

Next, based on the distribution of slope ratios found in our network (Figure S2 in Supplementary Data), we chose the slope ratio threshold of 0.15 and imposed directionality on the edges which had slope ratio less than 0.15. The resulting directed network is shown in Figure S3 and described in the Supplementary Data. While differential regulation of co-expressed genes belonging to the same pathway could be implemented in several ways, a remarkably consistent picture emerged when we looked at how this was operationally achieved at the sequence level in an organism. In bacterial and other prokaryotic systems, genes that encode for proteins necessary to perform coordinated functions in a particular pathway are typically clustered into a single transcriptional unit or operon that is transcribed into a single *polycistronic* mRNA coding for multiple proteins (De Hoon *et al.*, 2004). Indeed, we found a large number of instances of genes constituting an operon belonging to the same co-expressed cluster. For example, in cluster 2, genes could be grouped as *argC-argB-argD*, *carA-carB*, *argH-argG* and *rocF-rocD* according to their operon organization (Makarova *et al.*, 2001), the tryptophan metabolism genes of cluster 3 constitute the *trpEDCFB* operon (Du *et al.*, 2000), while the biotin biosynthesis genes of cluster 4 form the *bioWAFDBI* operon (Bower *et al.*, 1996; Perkins *et al.*, 1996). Overall, the fraction of operons preserved in our co-expression network (meaning that all genes of an operon belong to the same cluster) is found to be 82.6% with 38 of the 46 operons being preserved[1]. Moreover, we were able to uncover the sequence level basis for the differential regulation of genes in the same operon, namely the existence of alternative transcription termination sites as described next.

The network topology of one of the inferred sub-networks containing genes involved in biotin biosynthesis (Figure 5) implied that (*i*) the expression levels of all 6 genes were highly correlated across time and (*ii*) small changes in the expression level of *bioI* were translated into large changes in the expression of the remaining 5 genes (Figure 5B). The underlying biological reasons for these observations were found at the sequence level since the 6 genes are organized into an operon (Bower *et al.*, 1996; Perkins *et al.*, 1996) with two putative *rho*-independent transcription termination sites $t_1$ and $t_2$ (Figure 5C ) . This type of transcription termina-tion occurs when the transcribing RNA polymerase encounters a G-C rich region with dyad symmetry (inverted repeats). Once transcribed, the inverted repeats self-hybridize (anneal) to form a stem-loop structure which causes the RNA polymerase to pause. This allows the DNA strand to re-anneal and results in the release of the RNA polymerase along with the newly synthesized mRNA transcript. Consequently, the presence of the two termination sites suggested the existence of two transcripts of different lengths as shown in Figure 5C. This was indeed the case as previous experimental work (Perkins *et al.*, 1996) on the *bio* operon had uncovered a 7.2-kb RNA that corresponded to the entire 7-gene operon along with a 5.1-kb transcript that corresponded to only the first 5 genes. In addition to the difference in lengths, the relative amounts of the two transcripts were also determined to be very different. Specifically, the amount of the 5-gene transcript was found to be 8 times greater than that of the 7-gene transcript (Perkins *et al.*, 1996). This difference in relative abundance would translate into lower expression level of *bioI* (and *orf2*) as compared to the other 5 genes, in accordance with both our experimental measurements and the relationships implied by our network representation. From a biological control perspective, our results support previous suggestions (Perkins *et al.*, 1996) that the reaction catalyzed by the *bioI* protein product, which is the very first reaction in the linear biotin synthesis pathway, is the rate limiting step in this pathway.

---
[1] An important point to note is that operons and clusters are *not* equivalent in the sense that if the genes are known to be part of an operon then they can be expected to belong to the same cluster; however, all genes in a cluster do not constitute an operon.

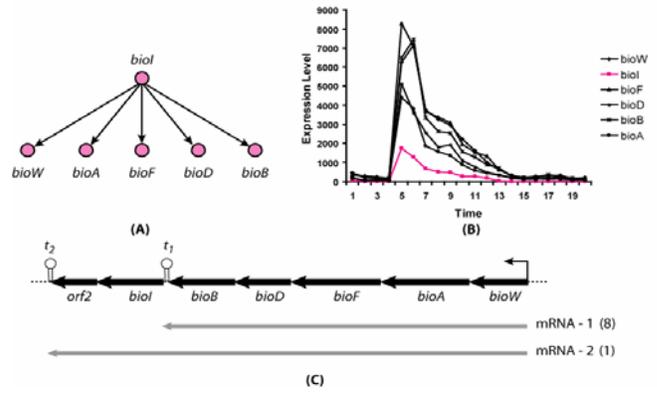

**Fig. 5.** (A) Sub-network inferred for the biotin biosynthesis genes (B) Measured expression profiles for the 6 genes (C) Sequence map of the B. subtilis bio operon (orf2: uncharacterized gene) and the reported relative abundances of the two transcripts (the shorter transcript was found to be 8 times more abundant than the longer one).

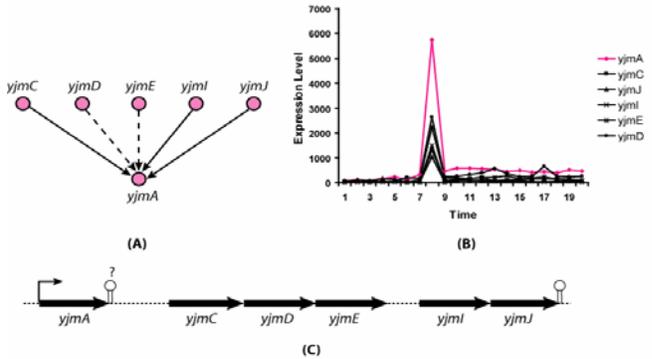

**Fig. 6.** (A) Sub-network inferred for the hexuronate utilization genes. (B) Measured expression profiles for the 6 genes (C) Sequence map of the B. subtilis *yjm* operon. The question mark indicates the computationally determined putative termination site.

Motivated by the results for the *bio* operon, we explored whether our approach could be used to validate/invalidate computational predictions of putative terminator sites. Typically, possible stem-loop structures between genes transcribed in the same direction are predicted by calculating the free-energy of base pairings in the stem region (de Hoon *et al.*, 2005). If the calculated free energy is negative and is below a specified threshold value, the sequence region is inferred to code for a putative transcription termination site. One such computationally predicted stem-loop site embedded within the *yjm* operon of *B. subtilis* is shown in Figure 6C (Rivolta *et al.*, 1998), along with the network inferred for some of the genes constituting this operon (Figure 6A) and their expression profiles in the cradle-to-grave data-set (Figure 6B). The directed arcs from *yjmD* and *yjmE* are included as dashed lines in Figure 6A because they barely miss the SR threshold of 0.15 with SR values of 0.214 (*yjmD* → *yjmA*) and 0.153 (*yjmE* → *yjmA*). In light of the results for the *bio* operon, the large difference in the expression level of *yjmA* as compared to the other genes suggests that the putative stem-loop structure is indeed a site for transcription termination. Other previously known instances of read-through terminators embedded within operons that we were able to identify included terminators in the *valS-folC* operon (read-through terminator after *valS*) and the *gap-pgk-tpi-pgm-eno* operon (read-through terminator after *gap*).

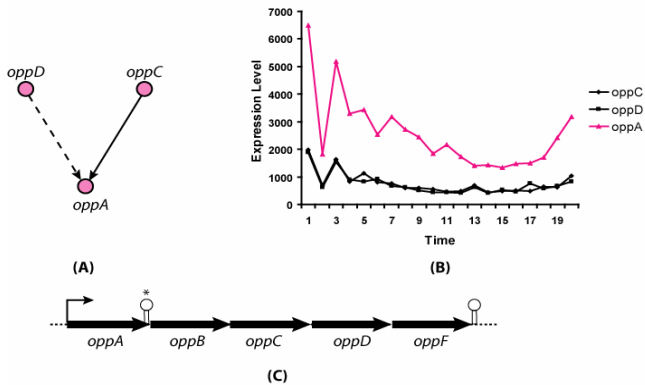

**Fig. 7.** (A) Sub-network inferred for the *opp* genes. The *oppD* → *oppA* arc is drawn as a dashed line because the R-square value between *oppD* and *oppA* is 0.845, which is just below the applied threshold of 0.90 (B) Measured expression profiles for the 3 genes (C) Sequence map of the B. subtilis opp operon with the newly identified putative intra-operon transcription termination site (indicated with a *) .

We are also able to predict the presence of truly novel putative intra-operon transcription termination sites as in the case of the *opp* operon shown in Figure 7. The genes *oppA* and *oppC* and *oppD* were found to comprise a 3 gene cluster with the directionality between them as shown in Figure 7A. These genes, along with *oppB*, *oppD* and *oppF*, constitute the *oppABCDF* operon encoding proteins involved in the oligopeptide transport system responsible for importing/exporting peptides of 3-5 amino acids across the cell membrane (Solomon *et al.*, 2003). The expression data for these genes (Figure 7B), in combination with their sequence map (Figure 7C), strongly suggest the presence of an intra-operon termination site between *oppA* and *oppB* since the expression level of *oppA* was much larger than all the other genes. Previously, using free-energy calculations, it had been hypothesized that the secondary structure of the RNA corresponding to the *oppA-oppB* intergenic region may enhance the stability of the mRNA molecule or function as a read-through terminator (de Hoon *et al.*, 2005). Our results support the read-through terminator hypothesis since there is large difference in the expression profiles of *oppA* and *oppCD*. This would not be expected if the sequence region was contributing to mRNA stability.

Next, we explore whether given the fact that we find good correspondence between our directed networks and the presence of intra-operon terminator sites, our analysis can be taken a step further to infer the relative strengths of known terminator stem-loop structures. Typically, two (or more) terminator sites are compared

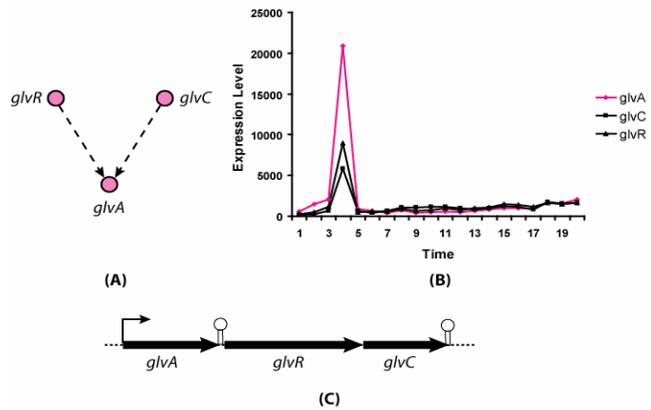

**Fig. 8.** (A) Implied sub-network for the *glv* genes involved in maltose metabolism. The arcs are shown with dashed lines since the edges barely miss the applied thresholds (B) Measured expression profiles for the 3 genes (C) Sequence map of the B. subtilis *glv* operon

based on their free-energy values. The more negative the free-energy, the more stable the stem-loop implying greater chance of termination at that site. Thus, a large difference in relative strengths of two terminators gets translated into significant difference in the amounts of the two possible transcripts, which corresponds to a strong directionality (indicated by a small slope ratio) in our analysis. One such operon of *B. subtilis* for which the free-energy values were available was the *glv* operon (Figure 8), where $\Delta G = -30.1$ Kcal/mol for the terminator after *glvA* and $\Delta G = -18.6$ Kcal/mol for the terminator after *glvC* (Yamamoto *et al.*, 2001). We find that the slope ratio of the *glvA-glvC* pair is 0.0678, suggesting that the terminator after *glvA* is much stronger than the one after *glvC*, in good agreement with the free energy results

In contrast to the *glv* operon, consider the pyrimidine biosynthesis genes (Figure 9). For this operon, previous studies predicted that the two terminator sites were approximately identical with respect to their transcription termination potential (Turner *et al.*, 1994; Lu *et al.*, 1995; Lu and Switzer, 1996). In particular, the free-energy values for the *pyrR-pyrP* and *pyrP-pyrB* terminator sites were reported to be $\Delta G = -15.3$ Kcal/mol and $\Delta G = -21.6$ Kcal/mol respectively (Turner *et al.*, 1994). As for the *glv* operon, our experi-mental data supported this claim since the relative distance between the expression profiles of *pyrR* and *pyrP* was approximately identical to that between *pyrP* and *pyrB*/*pyrC* and the slope ratio was found to be 0.9837 for the *pyrR-pyrP* pair and 0.8792 for the *pyrP-pyrB* pair.

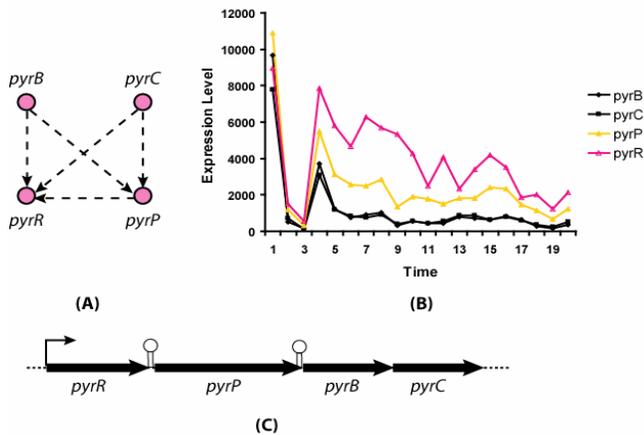

**Fig. 9.** (A) Implied sub-network for the pyrimidine biosynthesis genes. The arcs are shown as dashed lines since they miss the applied thresholds (B) Measured expression profiles for the 4 genes (C) Sequence map of the B. subtilis *pyr* operon

## 4 DISCUSSION

This paper presents a systematic framework for integrating network inference and topological analysis for uncovering biological knowledge from large-scale gene expression data. Our starting point was the gene co-expression network for *B. subtilis* as inferred from Affymetrix GeneChip® time-series data. We constructed the gene co-expression network of *B. subtilis* where each gene corresponds to a node and an un-directed edge exists between two nodes if the similarity (R-square value) between their expression profiles is above a threshold. One of the key contributions of our work is the formal estimation of this threshold value by applying ideas from network analysis. Specifically, we demonstrated how the clustering coefficient, which is a measure of local connectedness or cliquishness in a network, is the appropriate graph-theoretic metric that must be used for determining the cut-off value. This reliance on the clustering coefficient, as opposed to other topological measures such as degree distributions, path lengths, etc., was shown to be a natural consequence of the manifestation of the transitive property of linear functions at the network representation level. We expanded the scope of the gene co-expression network by proposing a novel measure, the *slope ratio*, as a means for imposing directionality on the edges. By doing so, we were able to account for both the temporal similarity of gene expression profiles as well as the similarity/dissimilarity in the absolute *levels* of gene expression. Both these attributes of the expression profiles are important as the similarity in temporal pattern is indicative of co-regulation of genes by an endogenous/exogenous signal while the difference in absolute levels provides information about the relative turnovers of the specific reactions in a biological pathway.

We found that the *B. subtilis* gene co-expression network was highly modular, consisting of one large connected sub-network and a large number of relatively smaller sub-networks. We observed no redundancy in our clustering results as there was no overlap between the expression signatures of the top six clusters. We verified that genes participating in the same biological pathway were co-expressed and consequently belonged to the same cluster and we identified a number of cases where some/all genes belonging to the same cluster constituted a single transcriptional unit or operon. Further probing of the operon structures of a number of gene clusters revealed that edge directionality, which was designed to capture the difference in expression levels, corresponded to the presence of intra-operon termination sites. We were able to (*i*) reconfirm previously experimentally determined termination sites as in the case of the *bio* operon, (*ii*) provide validation for computationally derived termination sites as in the case of the *yjm* operon, and (*iii*) quantify their relative strengths to complement free-energy based assessments. Given these results, we conclude that indeed there were important biological insights that were masked at the level of the un-directed co-expression network and that these were brought to light by considering the directed network. In the future, we plan to conduct a large-scale systematic study to see how well terminator free energy data, which is now available on a genome-wide scale for *B. subtilis* (de Hoon *et al.*, 2005), correlates with our slope ratio metric. Such a study would help identify instances where the two approaches (one based on sequence level information and one based on expression level information) support as well as complement each other.

## ACKNOWLEDGEMENTS

This work was partially supported by a Sloan Research Fellowship to R. A. Financial support by NSF award BES0120277 to C. D. M. is gratefully acknowledged. The authors would like to thank Dr. Jeffrey D. Varner from Genencor International for providing us with the microarray data used in this work and István Albert for helpful discussions.